\documentclass[fleqn,usenatbib]{mnras}

\usepackage{newtxtext,newtxmath}

\usepackage[T1]{fontenc}
\usepackage{ae,aecompl}

\usepackage{graphicx}	
\usepackage{amsmath}	
\usepackage{amssymb}	

\usepackage{todonotes}
\usepackage{xcolor}
\usepackage{multirow}
\usepackage{adjustbox}

\title[Multiple Populations during Cluster Formation]{On the Origin of Multiple Populations During Massive Star Cluster Formation}

\author[C.S. Howard et al.]{
Corey S. Howard,\thanks{E-mail: howardcs@mcmaster.ca}
Ralph E. Pudritz, 
Alison Sills, 
and William E. Harris
\\
Department of Physics \& Astronomy, McMaster University, 1280 Main Street West, Hamilton ON, L8S 4M1, CANADA\\
}

\date{Accepted 28 March 2019. Received 1 March 2019; in original form 11 December 2018}

\pubyear{2019}

\begin{document}
\label{firstpage}
\pagerange{\pageref{firstpage}--\pageref{lastpage}}
\maketitle

\begin{abstract}
We investigate the possibility that multiple populations in globular clusters arise as a natural by-product of massive star-cluster formation. We use 3D radiative hydrodynamics simulations for the formation of young massive clusters to track their chemical self-enrichment during their first 5 Myr. These clusters form embedded within filamentary Giant Molecular Clouds by a combination of gas accretion and rapid merging of protoclusters.  Chemical enrichment is a dynamic process happening as the young cluster assembles, so that the original (1P) and enriched (2P) subpopulations of stars form almost simultaneously.  Here we test two simple and opposite extremes for the injection of enriched material into the intracluster gas:  we assume either \emph{continuous injection} in a way that tracks the star formation rate; or \emph{sudden injection} by a single instantaneous event. Using helium abundance as a proxy for the enrichment, we find that realistic multiple population features can be reproduced by injecting a total helium mass amounting to a few percent of the cluster's total mass. The differences in individual growth histories can lead to widely differing 1P/2P outcomes.  These models suggest that dual or multiple populations can emerge rapidly in massive star clusters undergoing the typical mode of star cluster formation.
\end{abstract}




\section{Introduction}
Globular clusters are massive ($10^{4-7} M_{\odot}$) and low-metallicity ([Fe/H] $\lesssim 0$) star clusters that formed in the early universe \citep{Kruijssen} and are present in all large galaxies.
An outstanding puzzle is that many globular clusters host multiple populations of stars with distinct chemical abundance patterns \citep{Maclean2015,BastianLardo}.  The existence of multiple populations (MPs) in globular clusters (GCs) is nearly ubiquitous 
regardless of cluster metallicity and 
has been identified through abundance anomalies of proton capture 
elements such as C-N, Na-O, Mg-Al, and Na-F anticorrelations, as well as He 
abundance spreads \citep{Carretta2009,Piotto,Mucciarelli2014,Milone2018}. In many cases, GCs host 
two distinct stellar populations roughly in a 
1:1 ratio \citep{Caretta2009b} but with wide variety; 3 or more populations 
have been observed in several clusters, and in some cases the abundance distribution resembles a
roughly continuous spread with a few indentifiable clumps \citep{Milone2017}. 
MPs characterized by these abundance anomalies have not been seen in massive clusters younger than $\sim$2 Gyr \citep{BastianLardo}. 
	
MPs show abundance spreads primarily in light elements and rarely, for example, in Fe which is associated with Type II supernovae, suggesting that hot hydrogen burning is needed \citep{DeniHart}. Exotic origin scenarios for MPs have been proposed, but to date all encounter serious problems \citep{BastianLardo, Renzini, Bastian2015}. Several different concepts have attempted to explain the origin of MPs by invoking special conditions in the early universe or special processes \citep{DErcole2016,Renzini}. A number of scenarios assume that the enriched second population of stars (2P) forms out of material released by particular members of a first population (1P) such as AGB stars, 
massive stars, or binaries. However, these models have difficulty producing enough material to create the second populations (the ``mass budget" problem), often cannot reproduce the observed chemical abundance patterns in detail, and require a time lag of anywhere from $5 - 100$ Myr between populations, which is not supported by observations of either GCs or young massive clusters \citep{Nardiello,BastianLardo,Martocchia}.  To the limits of current measurements the 1P and 2P subpopulations have similar ages.

Before discussing stellar populations in GCs, we briefly summarize the main features of the current observations and theory of cluster formation.  

The birth sites of star clusters are now known to lie within overdense regions, known as clumps, within Giant Molecular Clouds (GMCs).  Observational studies of  star clusters across three decades in cluster mass ($10^3-10^5 M_{\odot}$) indicate that they have a continuum of physical properties that indicates a common formation mechanism \citep[see, e.g., the review of][]{krumholz2018}. The velocity dispersion of gas within GMCs is supersonic \citep{Larson1981}, usually interpreted as evidence for supersonic turbulence.  GMCs consist of networks of filamentary structures \citep[e.g.][]{Andre2014} and many simulations over the last two decades have shown that filaments are readily created  by gas compression that occurs at the intersection of shock waves in these supersonic turbulent conditions (eg. review \citep{MaclowKlessen}.  Stellar clusters are observed to form in the filaments, typically in clumps at the intersections (``hubs") of such systems of filaments \citep{Myers2009}. Moreover, gas accretion into forming star clusters occurs by observed filamentary flows \citep{Kirk2013}.  GCs form in the most massive GMCs since such clouds have more massive clumps \citep{HarrisPudritz1994, Reina-CamposKruijssen2017}.  As an example, \cite{Johnson2015} observed a clump within a massive GMC in the Antennae galaxies of the order of $5 \times 10^6 M_{\odot}$, in the right range to give birth to a young GC. 

Cluster formation is terminated by feedback, which for the most massive clusters, involves a variety of processes including radiative feedback (\cite{Dale2005, Murray2010}).  Our previous numerical simulations, which include radiative feedback effects of the forming clusters on their host GMCs, showed that there is a universal scaling relation between the maximum cluster mass in a GMC, and the GMC mass: $M_{max} \propto M_{GMC}^{0.78}$, across three decades of GMC mass \citep{NatAst}.   Thus, while earlier models of cluster formation viewed them as isolated entities each with their own peculiarities (open clusters, globular clusters, associations, etc), modern observations clearly indicate that they are parts of an extended hierarchical formation process that continues up to GMC scales and beyond.   It is within the context of this observationally grounded, physical picture of cluster formation that we can now begin to address the nature of the stellar populations in GCs. 

The observational fact that MPs are found in many or most GCs suggests that they may be a byproduct of this 
normal mode of star cluster formation for the most massive clusters.
The purpose of this paper is therefore to begin exploring whether or not 
MPs with realistic ranges of properties can plausibly emerge within this standard picture of cluster formation.
To build a quantitative description, we start with our 3D radiative hydrodynamics (RHD) simulations 
of massive star-cluster formation within 10$^7$ M$_{\odot}$ GMCs \citep{Howard2017-2,NatAst}. In these,
massive clusters grow rapidly (within $\simeq 5$ Myr) through an almost equal combination of rapid gas accretion from their natal GMC filaments, and merging with other protoclusters, and are certainly not isolated systems. The details of these radiation hydrodynamics simulations are laid out in \citet{NatAst}. 

These RHD simulations have not yet included any details of the chemistry or enrichment of the young stars inside the clusters.  Into these simulations, we therefore add one extra ingredient that will allow enrichment of the intracluster gas, but in a way that will not require a large temporal spread between 1P and 2P formation.  As an initial trial of such a model, we investigate two simple extreme cases.  The first mechanism assumes that enriched material is added continuously during cluster growth at a rate that tracks the star formation rate within the forming cluster.  This case will be referred to below as the \emph{continuous injection} (CI) alternative.
The second mechanism assumes an instantaneous single addition of enriched material, which is referred to below as \emph{sudden injection} (SI).
Most importantly, both mechanisms need to be active during the early stages of cluster formation before supernovae have cleared the intracluster gas, implying that the 1P and 2P stars form almost concurrently.  

Unlike previous scenarios, neither of these cases is viewed as happening within 
monolithic, isolated, single protoclusters; such objects are too idealized to represent real cluster formation and in any case lead to serious interpretive problems (see above).
Instead, we use simple post-processing techniques within our RHD simulations of GMCs to track the He abundance $Y$ of the stars and gas within their young star clusters, where $Y$ is used as a proxy for the overall level of chemical enrichment.  At this early stage of our modelling, the two opposite cases of enrichment rate that we calculate (CI and SI) 
are motivated by two general types of enrichment processes that have been discussed in the recent literature on He and various proton-capture elements that arise from the evolution of the most massive stars in clusters.  The two opposite extremes that we calculate, as noted above, are intended to bracket the current uncertainties in the theory of massive star formation and evolution (see Section 4 below for additional discussion of some of the  possibilities).

In section 2, a short review of the RHD simulations of giant GMCs is provided, followed by the details for our computation
of the internal enrichment of a model young massive cluster, under both of our extreme cases. In Section 3 the numerical
results of each case are shown, along with brief comparisons with observations. In Section 4, we suggest some possibilities for the types of massive stars responsible for the enrichment.   Finally, in Section 5 we give some additional discussion, a
summary of the method and its advantages, and prospects for future work.

\section{Method}
Our simulations of GMCs \citep{Howard2017-2,NatAst} were done with version 2.5 of the Adaptive Mesh Refinement (AMR) code \textsc{FLASH} \citep{Fryxell2000} which includes modules to treat hydrodynamics, self-gravity, radiative transfer, cooling processes, and star formation. Here, we provide only a brief summary of the most relevant modules here and direct the reader to our recent papers for more detail \citep{Howard2016,Howard2017-2}. Star clusters are represented by Lagrangian sink particles. When a region of gas in the GMC exceeds a specific density threshold and meets 
six other required criteria \citep{Federrath2010} that identify the particle as a region that will stay bound, a particle is created which 
can then interact with its surroundings gravitationally: as it moves through the simulation volume it can accrete gas 
within its radius, defined to be 2.5 cells at the highest level of refinement (typically about 1 parsec). 
The far smaller stellar scales are not resolved, which necessitates the use of a star formation subgrid model in the clusters. 

When the conditions for the formation of a cluster (sink particle) are first met, we consider its mass to be composed solely of gas that can be used for star formation as time goes on. 
This gas is converted to stars over time by randomly sampling a Chabrier IMF \citep{Chabrier2005} with an efficiency 
of 20\% of the unused gas mass per freefall time. The freefall time (t$_{ff}$) is calculated from the density formation threshold 
(see below). This efficiency is adopted to be consistent with local star-forming clumps \citep{Lada2003}. 
We sample the IMF 10 times per t$_{ff}$ to allow the cluster's stellar mass to increase smoothly. 
The masses of the stars formed in each cycle are recorded, and when more gas is added to the cluster through accretion, we add it to the reservoir of unused gas in the cluster. 

The luminosity $L(t)$ of each cluster is calculated directly from its stellar population at any time. Each star is treated as 
a blackbody, and metallicity-dependent analytic fits \citep{Tout1996} determine their main-sequence luminosity and 
temperature ($L, T_{eff}$)  We do not include protostellar evolution. The bolometric luminosity and the UV luminosity are passed to 
the radiative transfer scheme. We employ a hybrid-characteristics ray-tracing scheme \citep{Rijkhorst,Peters2010} to treat the 
radiative transfer and its associated feedback (see our previous papers for a detailed description \citep{Howard2016,Howard2017-2}). 
The propagation of both ionizing and non-ionizing radiation is followed and the DORIC routines \citep{Frank1994,Mellema2002} 
determine the ionization state of the gas.  We assume both the 
ionizing and non-ionizing opacities scale linearly with metallicity.  
We have also expanded the ray-tracing scheme to include the effects of single scattering radiation pressure induced by UV photons.

The main source of opacity in gas of solar and somewhat lower metallicity to both ionizing and non-ionizing radiation is due to dust grains.   It has been shown that the gas-to-``dust ratio scales linearly with the metallicity down to $0.1 Z_{\odot} $, and we adopt this scaling in our radiation transfer calculations \citep{NatAst}.     Below this value there is a deficit of dust with respect to gas that leads to a different form for the opacity-dust relation. 

To reduce the (large) computational time of the radiative transfer scheme, we employ a mass threshold of $10^4 M_{\odot}$
below which clusters are assumed not to emit radiation. 
These small clusters continue to form stars, accrete gas, and participate in gravitational interactions but they are 
not considered in the radiative transfer calculation. 
In any event, however, $>$90\% of the total luminosity in the simulation is contained in clusters above this mass.

In previous work, we adopted a density threshold for cluster formation of 10$^4$ cm$^{-3}$ because it represents the observational divide between 
starless and star-forming clumps in the local Milky Way \citep{Lada2003}. Several theoretical papers, however, argue that the threshold is instead environmentally dependent \citep{KrumMcKee,Padoan2011}. We therefore also examine simulations with higher thresholds of 10$^5$ and 10$^6$ cm$^{-3}$. Increasing the threshold density for cluster formation has two main effects:  
first, clusters will take longer to start forming because the gas needs to collapse to higher densities. 
Second, the star formation rates in the clusters are increased because we use a fixed star formation efficiency per freefall time, 
and t$_{ff}$ is shorter at higher densities. Nevertheless, increasing the threshold density negligibly affects the final mass of the most massive cluster \citep{NatAst}. 

Cluster sink particles are allowed to merge under the following conditions: they are separated by less than a particle radius (1.7 pc), their relative radial velocities are negative, and they are gravitationally bound to one another. When these conditions are met, all 
stellar and gas mass within them is transferred to the more massive particle, which is then repositioned at the system's center of mass.

From the entire suite of simulations of \citet{Howard2017-2,NatAst}, for our present purpose we select the ones for 10$^7$ M$_{\odot}$ supergiant GMCs with metallicity 0.1 Z$_{\odot}$ in order to match a typical GC. In these models the GMC is initially spherical with a radius of 77 pc and is embedded in a cubic box 173 pc across. The smallest resolution of the simulation grid is 0.6 pc and outflow boundary conditions are used for the domain edges. The total mass in the simulation volume is therefore not conserved as matter can flow out of the box. The initial density profile is uniform in the central half of the cloud and decreases as r$^{-3/2}$ in the outer half, with a quadratic fit applied at the transition region to ensure a continuous and smooth density distribution. The density outside the GMC is 100 times less than the density at the cloud surface and its temperature is chosen such that the GMC and external medium are in pressure equilibrium. The temperature inside the cloud is initially 10 K.

We note that GC formation at redshifts 5 - 7 would mean that then the CMB temperature would be 15-20 K, setting a `floor' to the initial temperature warmer than 10 K. This will affect the fragmentation scale of the dense molecular gas, but will not change the relative numbers of massive stars in clusters that are the main drivers of the cluster radiation fields.  Therefore, we retain the 10 K temperatures of GMCs appropriate to the local universe. 

Each GMC is overlaid with a Burgers turbulent velocity spectrum, as in \citet{Girichidis2011}. The turbulent spectrum contains a natural (random) mixture of solenoidal and compressive modes, and creates the highly filamentary structure in which clusters eventually form.  The turbulence is not driven as the simulation evolves. The strength of the turbulence is determined by choosing the initial virial parameter 
$\alpha = 2 E_{kin} / |E_{grav}|$. We set $\alpha_0 =3$ (i.e.~ initially unbound) because, as shown in \citet{Howard2016}, it results in low SFEs to match observations, but the cloud quickly becomes virialized since the turbulence is not continously driven. The same velocity spectrum is applied to each cloud in order to isolate the effects of radiative feedback and metallicity.

Each model is evolved for $\sim$5 Myr. The simulations are ended at this time for two reasons. Firstly, the computational expense increases with $t$ because the gas collapses to higher densities, and the number of radiating clusters grows with time. Secondly, we do not include the effects of supernovae so we stop the simulations before the supernova phase can significantly alter the system's evolution \citep{NatAst}. Nevertheless, by this stage the 1P/2P enrichment patterns for the cluster stars have 
already been set (see below); the supernova stage will primarily affect the dynamical evolution of the cluster, and the fraction of its initial mass that will survive as a bound cluster \citep{BailinHarris}.

\subsection{Enrichment Calculation} 

The cluster subgrid model described above outputs each cluster's properties as a function of time. Most importantly, this includes 
the total cluster mass; its reservoir gas mass (i.e.~gas that has been incorporated into the cluster particle but has not yet been used for star formation); 
and the masses and formation times of the individual stars formed in each cluster.  Using this information, post-processing (described below) 
is used to track the He abundance ($Y$) in both the gas and the stars inside the clusters. 

\begin{figure}
\includegraphics[width=0.9\linewidth]{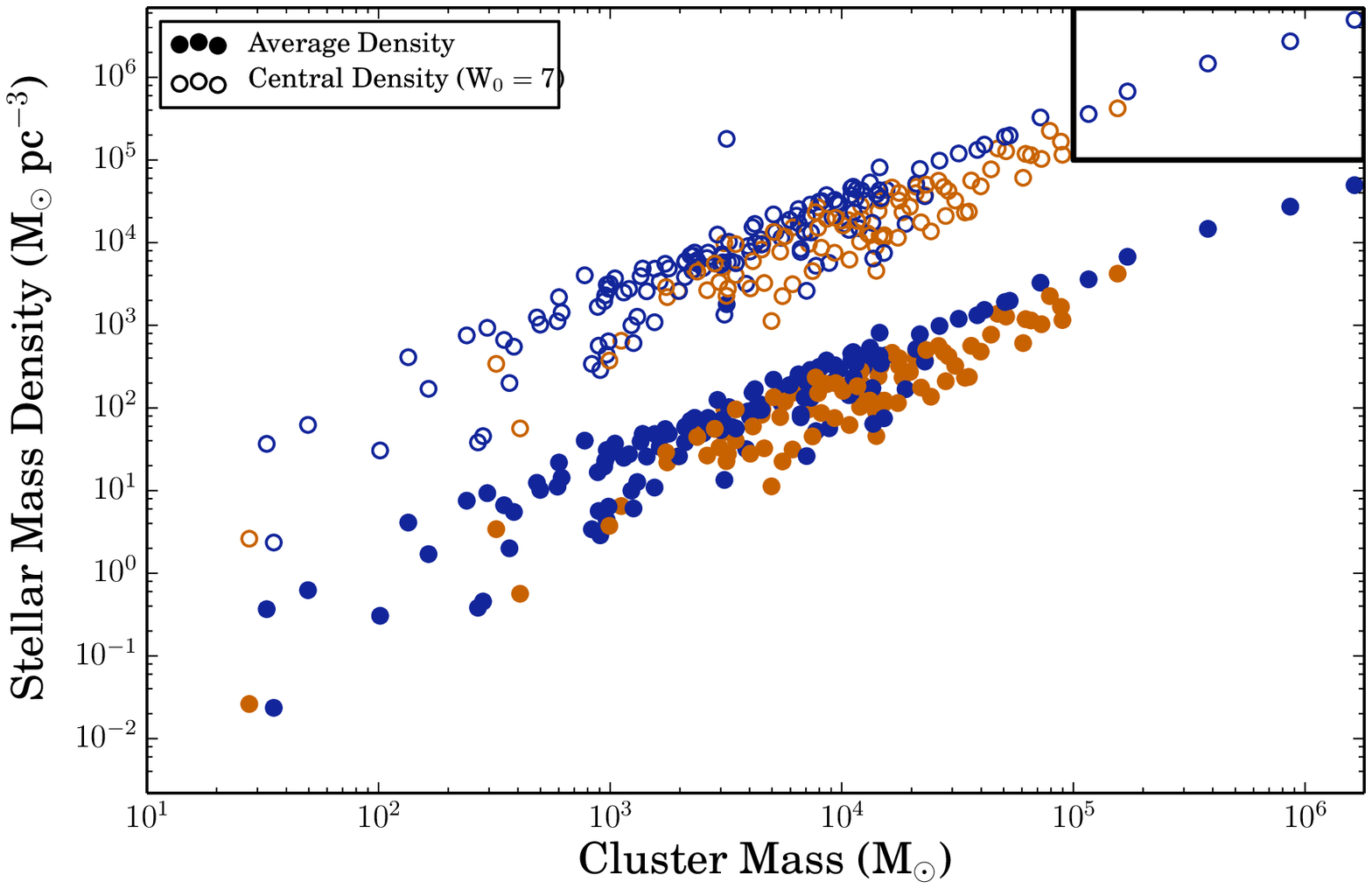}
\includegraphics[width=0.9\linewidth]{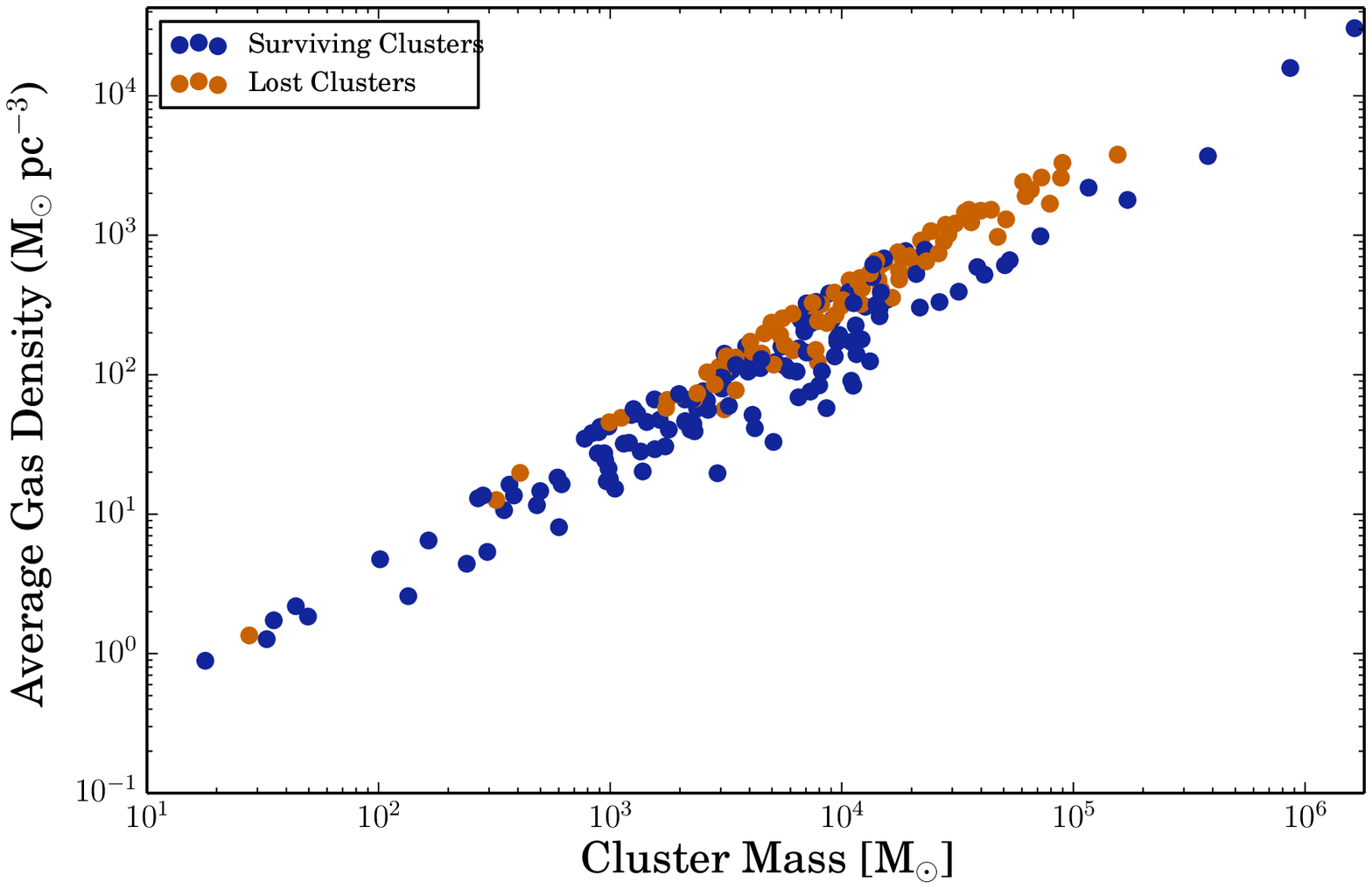}
\caption{\textbf{Stellar and gas densities in simulated clusters.} 
	\emph{Top:} The final average stellar densities (filled  circles -- lower sequence) and the central stellar densities assuming W$_0$ $=$ 7 (open circles -- upper sequence) for the simulation with a formation density threshold of 10$^4$ cm$^{-3}$. Blue markers represent clusters that survive to the end of the simulation while orange circles represent clusters that either merged to a larger cluster or left the simulation volume. The black box represents the area of interest after applying our mass and density cuts.  Sparse and low-mass clusters will be destroyed by two-body relaxation, tidal stripping, and interactions with nearby molecular clouds. \emph{Bottom:} The final average gas density within 
the clusters.}
\label{fig:densities}
\end{figure}

In Figure \ref{fig:densities}, we show the average and central densities for the star clusters that survive until the end of the simulation (blue) or are removed either through merging or leaving the simulation volume (orange). 
The outlined (in black) box demarcates the small subset of clusters for which we complete the enrichment calculations:  these are deliberately the densest and most massive ones.   
There are 13 model clusters in total that are the most likely candidates for young GCs:  i.e.~those with final masses above 10$^5$ M$_{\odot}$ and 
central stellar densities above 10$^5$ cm$^{-3}$.
Since we do not resolve the stellar distribution within the clusters, the central density cut assumes the stars follow a King profile \citep{King} with W$_0$ $=$ 7 which is typical for GCs \citep{Miocchi2013}.   

The initial pristine He abundance in the GMC is set at $Y_0$ $=$ 0.247 \citep{Planck}. When a cluster first forms, its mass is taken to be 
all gaseous and is assigned a He abundance of $Y_0$. 
Stars are then formed and He is added to the gas reservoir of the cluster through one of the 
two enrichment mechanisms described below (CI or SI), raising the He abundance above $Y_0$. When stars form, they are assigned the instantaneous 
$Y$ value of the reservoir and any He used in their formation is removed from the gas. The $Y$ values of the stars are recorded so that stellar abundance distributions can be studied once the model has evolved for the length of the simulation. 

Both \emph{gas accretion} and \emph{mergers with other clusters} can affect the internal $Y$ of the massive young cluster. We assume that any gas accreted directly from the host GMC has an abundance of $Y_0$, so strong gas accretion can therefore lower $Y$ if it has been enriched above $Y_0$. By assuming that the intercluster GMC gas always remains at $Y_0$, and that there is no mass loss from the cluster particles over these first few Myr of evolution, we are also implicitly assuming that the He and metals produced by stellar processes remain bound to their host cluster. Previous energy-based calculations on the retention of metals released by supernovae in massive star clusters indicate this is a reasonable approximation \citep{BailinHarris}; importantly, it would not be valid for lower-mass clusters. When a cluster merger occurs, the gas reservoir and the stellar population of the two clusters are combined.  Lastly, we also complete the above enrichment calculation for the merging partner before combining their gas and stellar populations. 

\subsection{Continuous Enrichment}

In the continuous injection (CI) picture, He is added continously throughout the evolution 
at a rate that matches the overall star formation rate within small statistical fluctuations.
To make the modelling specific, we recognize that the stellar abundance patterns observed in MPs, if they are produced in only the first few Myr of cluster evolution, should result from the most massive stars in the cluster, the most prominent examples of which are O-star binaries (OSBs, described more completely in Section 4 below).  The CI case,
in our simple approach, has two free parameters. The first is the stellar age at which mass loss occurs, 
for which we adopt $t(onset) = 1$ Myr, allowing ample time for enrichment to occur over the 5 Myr 
timescale of our simulation. We note that there is no accepted time for the onset of mass loss 
from OSBs but it certainly occurs in $<$5 Myr \citep{Petrovic2005} and likely depends on the binary system masses, orbital separation, and eccentricity. 

The second parameter in this model is the mass fraction $f_{He}$ of these stars that is returned to the intracluster gas as He. We choose 3 sample values for this parameter: 10\%, 25\%, and 50\%. Stellar evolution models of massive-star binaries demonstrate that the average $Y$ in these objects is 0.34 with extreme values reaching as high as 0.64 \citep{deMink2009}. This, combined with dynamical models that find only a small percentage of mass ($\sim$10\%) transferred between stars is actually retained \citep{Petrovic2005}, provides support for our adopted values. 

In this CI picture, the enrichment directly traces the formation rate of massive stars.  This rate 
is determined in part by the density formation threshold; in part by the availability of gas; and in part by the number of mergers of smaller clusters to form the larger final one. Therefore, the total enrichment in different individual clusters spans a large range of values, and does not have a simple dependence on any of those parameters.

\subsection{Sudden Enrichment}

The sudden injection (SI) picture is essentially the opposite of CI.  In SI, a large amount of enriched gas is injected into the forming cluster in a single delta-function event
at one particular time.  This is clearly an artificial assumption, but it is chosen because it is the natural physical and logical extreme opposite to the CI case. In a sense SI is simpler to calculate than CI, because the only free parameters are the total amount of He added to the intracluster gas relative to the cluster mass, $M_{He}/M_{cl}$, and the particular time of the event.
The Y-abundance of the intracluster gas suddenly jumps upward at the injection time, but \emph{after} the enrichment occurs, Y(gas) will gradually decrease again as more gas (with its pristine abundance) continues to
flow into the cluster.  This effect is the opposite of the gradually increasing Y(gas) produced
by the CI model, and can thus produce quite different final abundance patterns at the end of
the modelling run.

As will be seen in the Results below, an extremely important feature of these models that 
emerges automatically is that a wide variety of abundance distribution patterns (MPs) is
produced \emph{even within the same mechanism} (CI or SI).  This variety is directly due
to the individual growth histories of the model clusters, which can strongly differ in their
gas accretion and merger rates.  These differences, in turn, are expected for forming clusters in dynamically evolving filaments within turbulent GMCs.  In a much simpler cluster formation
scenario (such as isolated monolithic collapse) this cluster-to-cluster variance would have to be inserted by hand.  The major advantage of our models is that we start by already knowing the gas inflow and cluster merger rates as a function of time.

\section{Results}

As noted above, we follow the He enrichment histories of 13 model clusters with final masses that exceed 10$^5$ M$_{\odot}$ and with central stellar densities above 10$^5$ $M_{\odot} pc^{-3}$.  These also cover our three different assumed threshold densities for cluster formation: 10$^4$ cm$^{-3}$ (similar to local molecular clouds), 10$^5$ cm$^{-3}$, or 10$^6$ cm$^{-3}$. 

For each mechanism (CI or SI), our simple enrichment model is characterized by the amount of He added to the intracluster gas, and the time(s) since cluster formation when He is added.  In SI the enriched gas is added all at once, whereas in CI 
it is added through a long series of small events following the massive-star formation rate.  
In both cases, we do not have any strong initial constraints on the parameters to work with, 
so we run a suite of models that cast a rather wide net across parameter space, with the purpose of isolating the much smaller region of that space that will best match the available observations.  We expect, therefore, that most of our trials will ``miss'' the target region with enrichments that are too high or too low, and can be ruled out.

\begin{figure}
\includegraphics[width=0.9\linewidth]{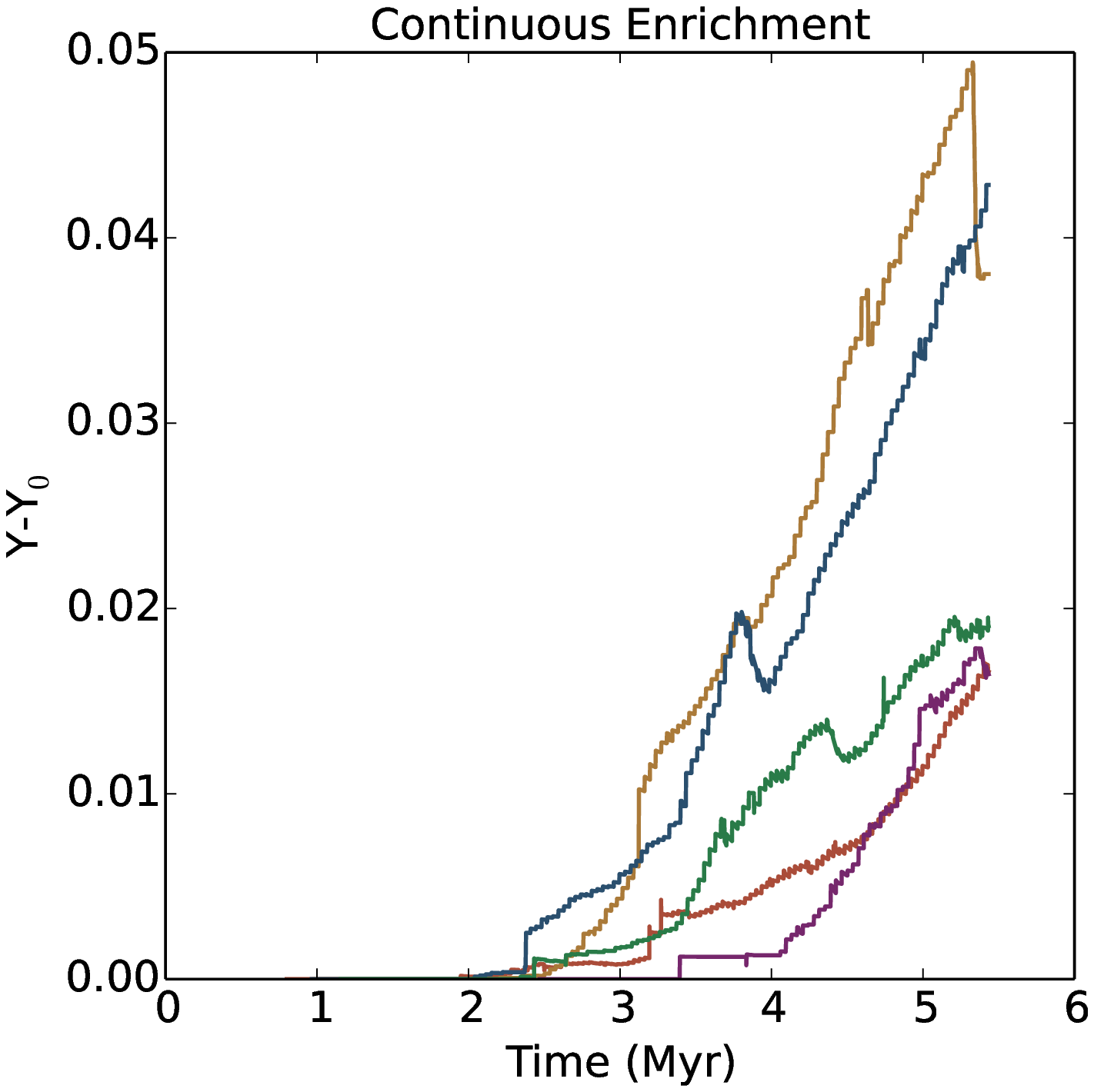}\\
\includegraphics[width=0.9\linewidth]{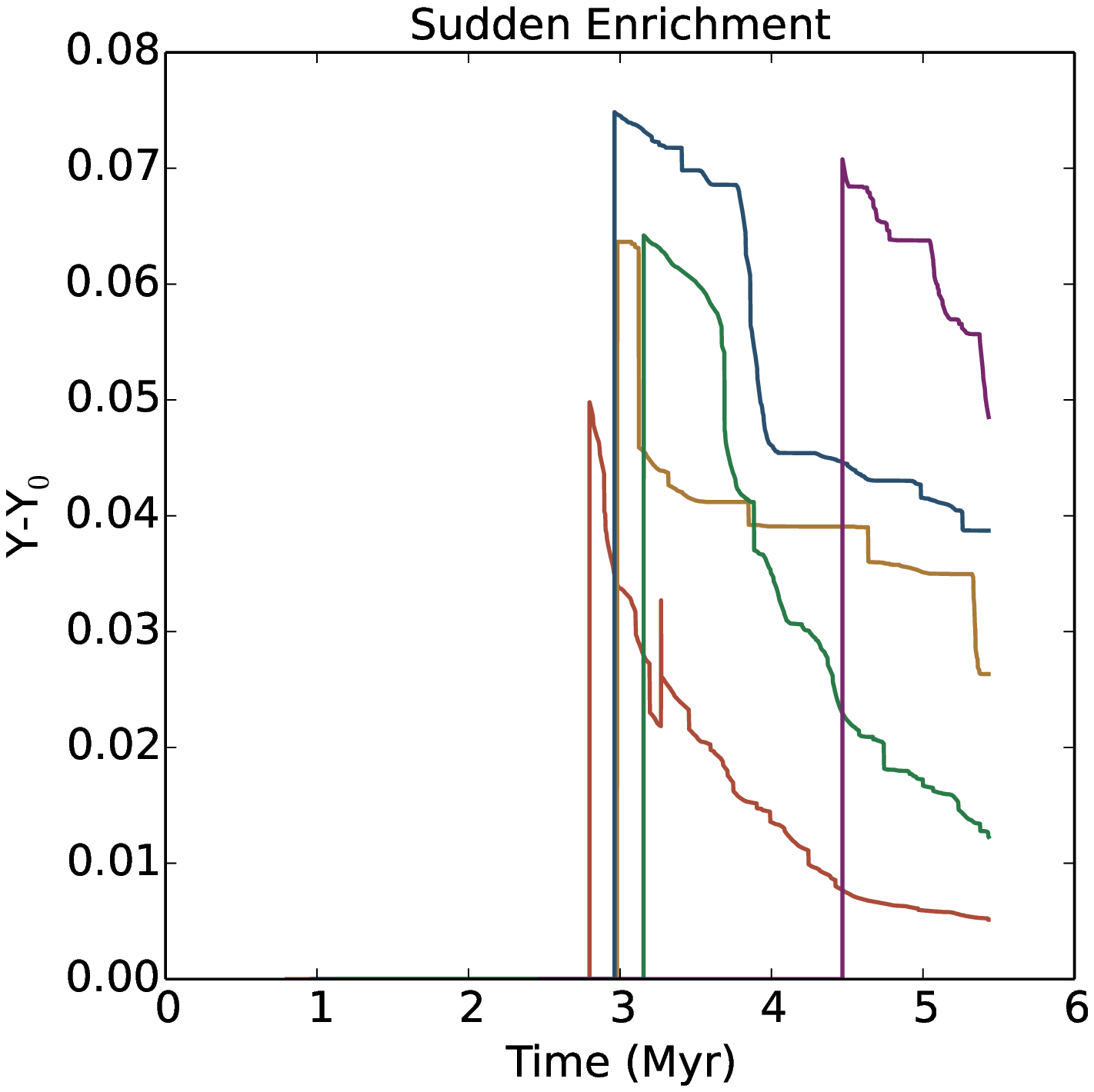}\\
\caption{\textbf{He gas abundance versus time for two different models.} The intracluster gas He abundance ($Y$) for the clusters that exceed the mass and stellar density cuts described 
in text. \emph{Top panel:} enrichment versus time in the CI model is shown 
for 5 model clusters with $f_{He} = 0.25$. 
The delay between the birth of the cluster and the onset of enrichment 
happens because the cluster must grow to sufficient mass to form massive stars 
that eventually shed their mass $> 1$ Myr after their formation.  
\emph{Bottom panel:} The enrichment tracks for the SI model are shown for 5 clusters with M$_{He}$/M$_{cl}$ $=$ 0.05. 
A sudden, single injection of He happens 2 Myr after the start of star formation.}
\label{fig:enrichmenttime}
\end{figure}

Typical examples of $Y$(gas) versus time are shown in Figure \ref{fig:enrichmenttime}, corresponding to ``midrange'' values of the mass ratios $f_{He} = 0.25$ for CI, and M$_{He}$/M$_{cl}$ $=$ 0.05 for SI. Note that for the CI case, 
$Y$ gradually increases as more stars are formed and the most massive ones release new He, 
but even here $Y(t)$ does not always increase monotonically, because pristine gas accretion and mergers with less enriched clusters can decrease $Y$ again, at least temporarily. 
By contrast, as mentioned above the SI case experiences an instantaneous increase in $Y$, but since there is no further injection of He and pristine gas accretion is still ongoing, $Y$ typically decreases after this point unless a merger occurs with an enriched cluster. 

We present two diagnostics to characterize the resulting stellar He abundances. 
The first is the total He abundance spread $\Delta Y = Y_{max} - Y_0$, the maximally enriched level minus the pristine value. In general, 
$\Delta Y$ does not exceed 0.1 in observed cluster, and current limits of detectability are 
$\Delta Y \simeq 0.01 - 0.03$ \citep{WK2016, Deni2017, Prantzos, Milone2018}. Our second diagnostic is the number 
ratio of first to second population stars (1P/2P) where the divide between populations is defined as $\Delta Y$ $=$ 0.01; i.e. essentially any enrichment above $Y_0$ is taken to belong to 2P. In Tables 1 and 
2, we include further information regarding the model clusters and their stellar abundance spreads 
for the simulation with the baseline $10^4$ cm$^{-3}$ formation density threshold.  
The tabulations include cluster masses, total He mass injected into the gas, and the mean and median $Y$ values for the stars. We also include the lower and upper quartiles for the stellar abundances.

\begin{table*}
\centering
\begin{adjustbox}{width=1\textwidth}
\small
\begin{tabular}{|c|c|c|c|c|c|c|c|c|}
\hline
Stellar He Fraction ($f_{He}$) & Final Total Mass (M$_{\odot}$) & Final Stellar Mass (M$_{\odot}$) & Total He Added (M$_{\odot}$) & $\Delta$Y & Mean Y & Median Y & Lower Quartile-Upper Quartile & Population Ratio (1P/2P) \\ \hline
\multirow{5}{*}{0.10} & 1.64$\times$10$^6$ & 1.02$\times$10$^6$ & 8.90$\times$10$^3$ & 0.253 & 0.249 & 0.249 & 0.248-0.251 & 1253 \\
                      		   & 8.64$\times$10$^5$ & 5.46$\times$10$^5$ & 5.55$\times$10$^3$ & 0.014 & 0.250 & 0.250 & 0.248-0.252 & 107 \\
				   & 3.81$\times$10$^5$ & 3.07$\times$10$^5$ & 3.43$\times$10$^3$ & 0.017 & 0.251 & 0.249 & 0.247-0.254 & 5.8 \\
				   & 1.71$\times$10$^5$ & 1.35$\times$10$^5$ & 1.62$\times$10$^3$ & 0.020 & 0.252 & 0.249 & 0.247-0.255 & 4.7 \\
				   & 1.16$\times$10$^5$ & 7.22$\times$10$^4$ & 5.53$\times$10$^2$ & 0.007 & 0.249 & 0.248 & 0.247-0.250 & N/A \\ \hline

\multirow{5}{*}{0.25} & 1.64$\times$10$^6$ & 1.02$\times$10$^6$ & 2.26$\times$10$^4$ & 0.280 & 0.253 & 0.252 & 0.249-0.256 & 1.9 \\
                                   & 8.64$\times$10$^5$ & 5.46$\times$10$^5$ & 1.39$\times$10$^4$ & 0.034 & 0.255 & 0.254 & 0.248-0.260 & 1.3 \\
				   & 3.81$\times$10$^5$ & 3.07$\times$10$^5$ & 8.58$\times$10$^3$ & 0.040 & 0.257 & 0.252 & 0.247-0.264 & 1.8 \\
				   & 1.71$\times$10$^5$ & 1.35$\times$10$^5$ & 4.04$\times$10$^3$ & 0.048 & 0.259 & 0.252 & 0.247-0.266 & 1.1 \\
				   & 1.16$\times$10$^5$ & 7.22$\times$10$^4$ & 1.38$\times$10$^3$ & 0.018 & 0.251 & 0.248 & 0.247-0.255 & 4.2 \\ \hline

\multirow{5}{*}{0.50} & 1.64$\times$10$^6$ & 1.02$\times$10$^6$ & 4.45$\times$10$^4$ & 0.306 & 0.259 & 0.258 & 0.514-0.265 & 0.87 \\
                      		   & 8.64$\times$10$^5$ & 5.46$\times$10$^5$ & 2.77$\times$10$^4$ & 0.065 & 0.263 & 0.262 & 0.250-0.273 & 0.83 \\
				   & 3.81$\times$10$^5$ & 3.07$\times$10$^5$ & 1.72$\times$10$^4$ & 0.076 & 0.266 & 0.257 & 0.247-0.280 & 1.1 \\
				   & 1.71$\times$10$^5$ & 1.35$\times$10$^5$ & 8.08$\times$10$^3$ & 0.090 & 0.270 & 0.258 & 0.247-0.283 & 0.97 \\
				   & 1.16$\times$10$^5$ & 7.22$\times$10$^4$ & 2.77$\times$10$^3$ & 0.033 & 0.256 & 0.250 & 0.247-0.262 & 2.1 \\ \hline
\end{tabular}
\end{adjustbox}
\caption{Cluster properties and stellar abundance statistics for the CI enrichment model in the simulation with a cluster density formation threshold of 10$^4$ cm$^{-3}$.}
\end{table*}

\begin{table*}
\centering
\begin{adjustbox}{width=1\textwidth}
\small
\begin{tabular}{|c|c|c|c|c|c|c|c|c|}
\hline
 Cluster He Fraction (M$_{He}$/M$_{cl}$) & Final Total Mass (M$_{\odot}$) & Final Stellar Mass (M$_{\odot}$) & Total He Added (M$_{\odot}$) & $\Delta$Y & Mean Y & Median Y & Lower Quartile-Upper Quartile & Population Ratio (1P/2P) \\ \hline
\multirow{5}{*}{0.01} & 1.64$\times$10$^6$ & 1.02$\times$10$^6$ & 2.24$\times$10$^3$ & 0.014 & 0.249 & 0.248 & 0.248-0.250 & 49 \\
                      & 8.64$\times$10$^5$ & 5.46$\times$10$^5$ & 3.38$\times$10$^3$ & 0.014 & 0.250 & 0.250 & 0.247-0.252 & 11 \\
                      & 3.81$\times$10$^5$ & 3.07$\times$10$^5$ & 2.89$\times$10$^3$ & 0.016 & 0.252 & 0.247 & 0.247-0.256 & 4.0 \\
		      & 1.71$\times$10$^5$ & 1.35$\times$10$^5$ & 9.23$\times$10$^2$ & 0.016 & 0.251 & 0.247 & 0.247-0.255 & 33 \\
                      & 1.16$\times$10$^5$ &7.22$\times$10$^4$  & 9.81$\times$10$^2$ & 0.015 & 0.251 & 0.247 & 0.247-0.259 & 2.3 \\ \hline

\multirow{5}{*}{0.05} & 1.64$\times$10$^6$ & 1.02$\times$10$^6$ & 1.12$\times$10$^4$ & 0.065 & 0.257 & 0.254 & 0.252-0.261 & 2.2 \\
                      & 8.64$\times$10$^5$ & 5.46$\times$10$^5$ & 1.69$\times$10$^4$ & 0.063 & 0.263 & 0.261 & 0.247-0.273 & 0.84 \\
                      & 3.81$\times$10$^5$ & 3.07$\times$10$^5$ & 1.44$\times$10$^4$ & 0.073 & 0.271 & 0.247 & 0.247-0.291 & 1.2 \\
		      & 1.71$\times$10$^5$ & 1.35$\times$10$^5$ & 4.62$\times$10$^3$ & 0.063 & 0.265 & 0.247 & 0.247-0.285 & 1.3 \\
                      & 1.16$\times$10$^5$ &7.22$\times$10$^4$  & 4.90$\times$10$^3$ & 0.067 & 0.266 & 0.247 & 0.247-0.302 & 2.3 \\ \hline

\multirow{5}{*}{0.1}  & 1.64$\times$10$^6$ & 1.02$\times$10$^6$ & 2.24$\times$10$^4$ & 0.119 & 0.267 & 0.260 & 0.257-0.273 & 0.33 \\
                      & 8.64$\times$10$^5$ & 5.46$\times$10$^5$ & 3.38$\times$10$^4$ & 0.115 & 0.277 & 0.274 & 0.247-0.298 & 0.84 \\
                      & 3.81$\times$10$^5$ & 3.07$\times$10$^5$ & 2.89$\times$10$^4$ & 0.133 & 0.293 & 0.247 & 0.247-0.332 & 1.2 \\
		      & 1.71$\times$10$^5$ & 1.35$\times$10$^5$ & 9.23$\times$10$^3$ & 0.115 & 0.281 & 0.247 & 0.247-0.320 & 1.2 \\
                      & 1.16$\times$10$^5$ &7.22$\times$10$^4$  & 9.81$\times$10$^3$ & 0.124 & 0.281 & 0.247 & 0.247-0.350 & 2.3 \\ \hline
\end{tabular}
\end{adjustbox}
\caption{Cluster properties and stellar abundance statistics for the SI enrichment model in the simulation with a cluster density formation threshold of 10$^4$ cm$^{-3}$.}
\end{table*}

\begin{figure}
\includegraphics[width=0.9\linewidth]{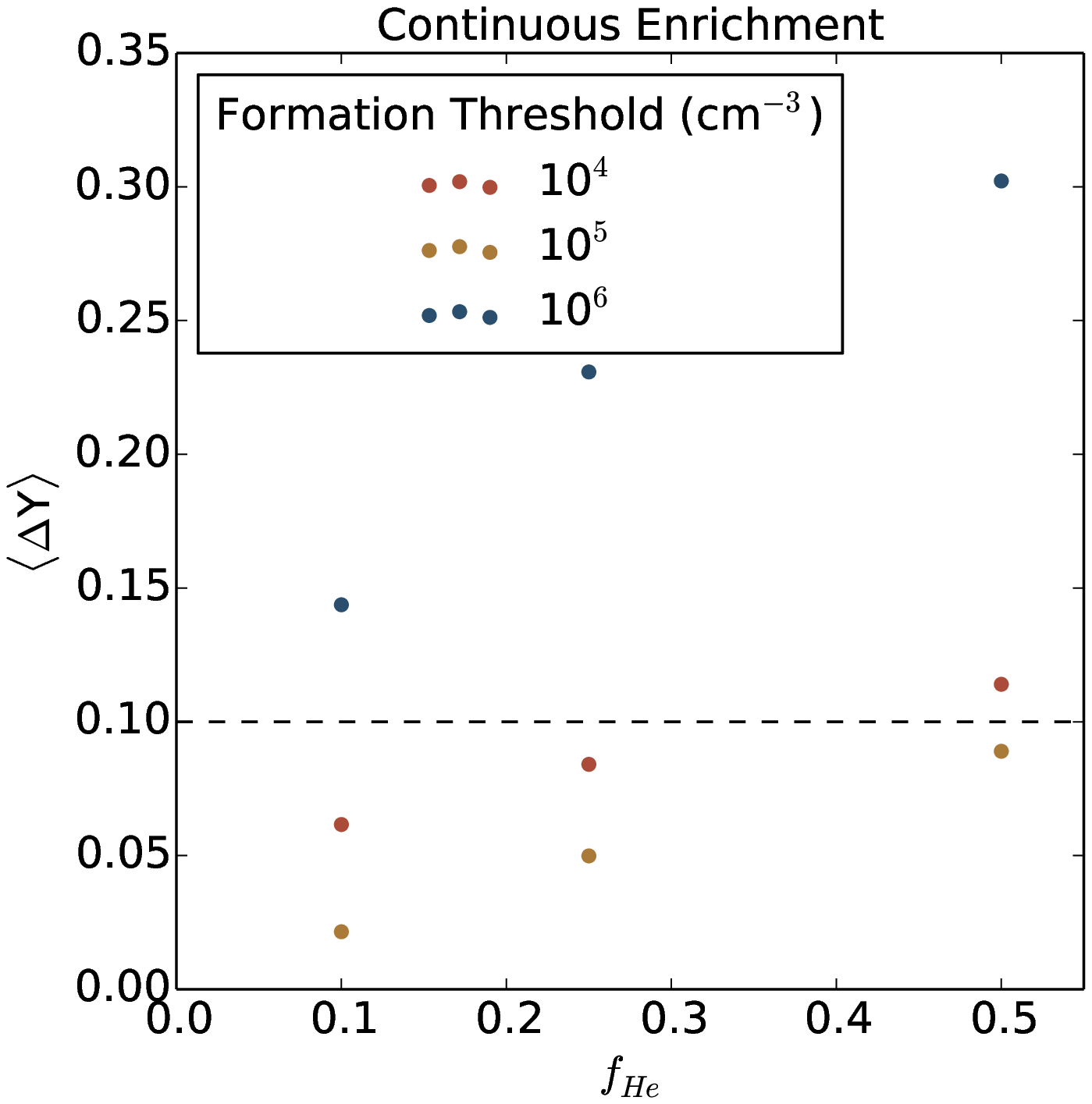} 
\includegraphics[width=0.9\linewidth]{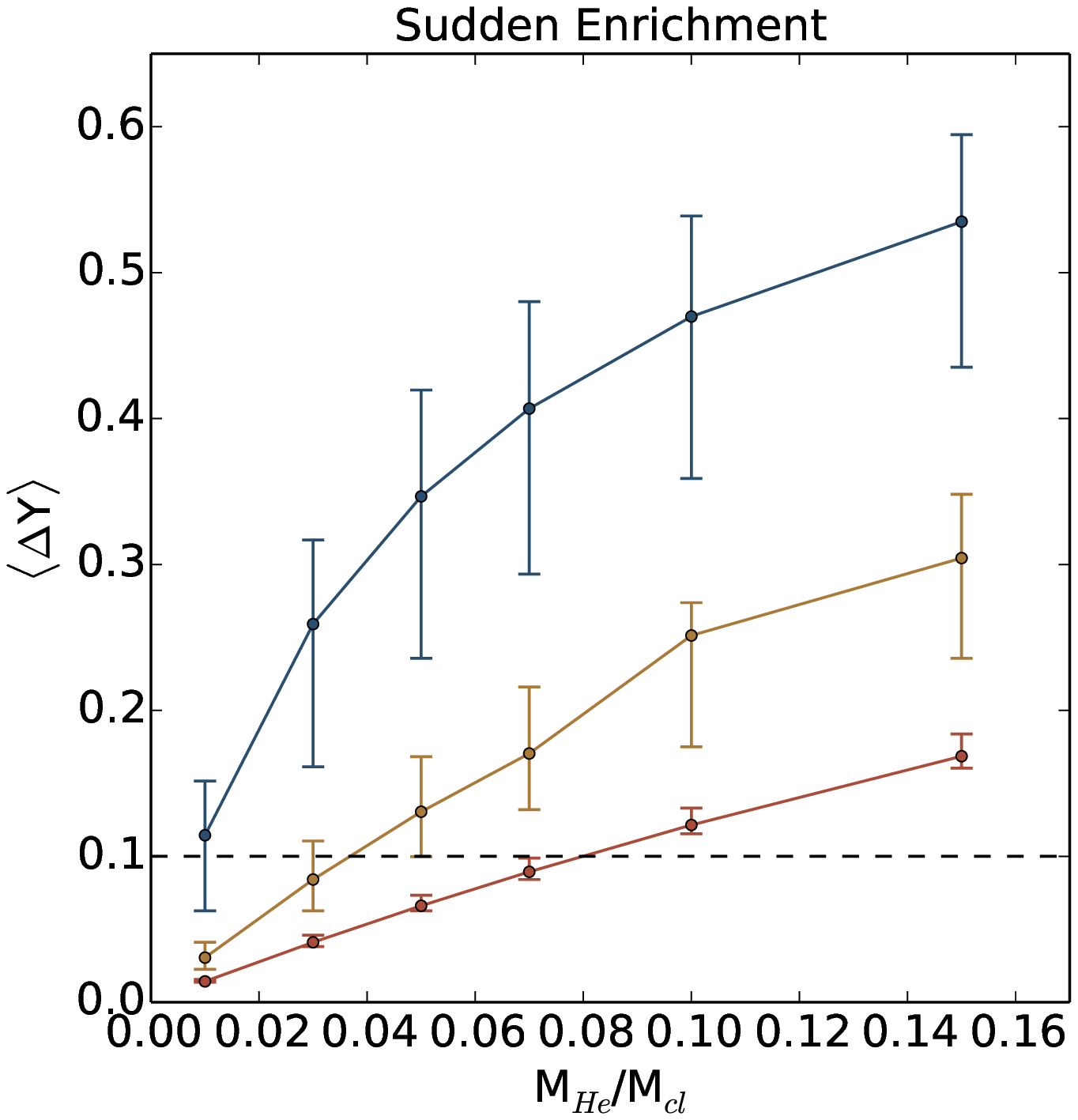} 
\caption{\textbf{Average Stellar Abundance Spreads for two different models.} 
	The average $\Delta Y$ as a function of $f_{He}$ for the CI case 
	(top) and M$_{He}$/M$_{cl}$
for the SI case (bottom) for various cluster formation density thresholds. 
Each point represents the average for all clusters exceeding our adopted mass and density cuts. 
The error bars on the SI plot span the total spread of $\Delta Y$ across clusters. 
Error bars are not plotted for the CI model since the spread exceeds 
the axis limits. The horizontal dashed line represents an approximate observational limit to $\Delta Y$
\citep{Milone2018}.}
\label{fig:Abundances}
\end{figure}

In Figure \ref{fig:Abundances}, $\Delta Y$ is shown as a function of $f_{He}$ 
(bottom panel) or M$_{He}$/M$_{cl}$ (top panel) for the different
formation density thresholds. Each point represents the average $\langle \Delta Y \rangle$ of the clusters. 
Observational results show that the majority of real
clusters fall within $\Delta Y < 0.05$ and rarely exceed 0.1 \citep{Milone2018}. 
Thus in both of our model cases, a threshold of 10$^6$ cm$^{-3}$ results in an excessively large $\Delta Y$ compared to observations. These clusters tend to have a low gas to stellar mass ratio (M$_{gas}$/M$_{*}$) due to the higher SFRs associated with high threshold values. The lower threshold values (characteristic of cluster formation in galactic GMCs), however, do have abundance spreads $\Delta Y < 0.1$ consistent with observations. For the CI case, most results produce $\langle \Delta Y \rangle$ $<$ 0.1, but for SI only M$_{He}$/M$_{cl}$ $\lesssim$ 0.07 produces realistic results.

\begin{figure}
\includegraphics[width=.9\linewidth]{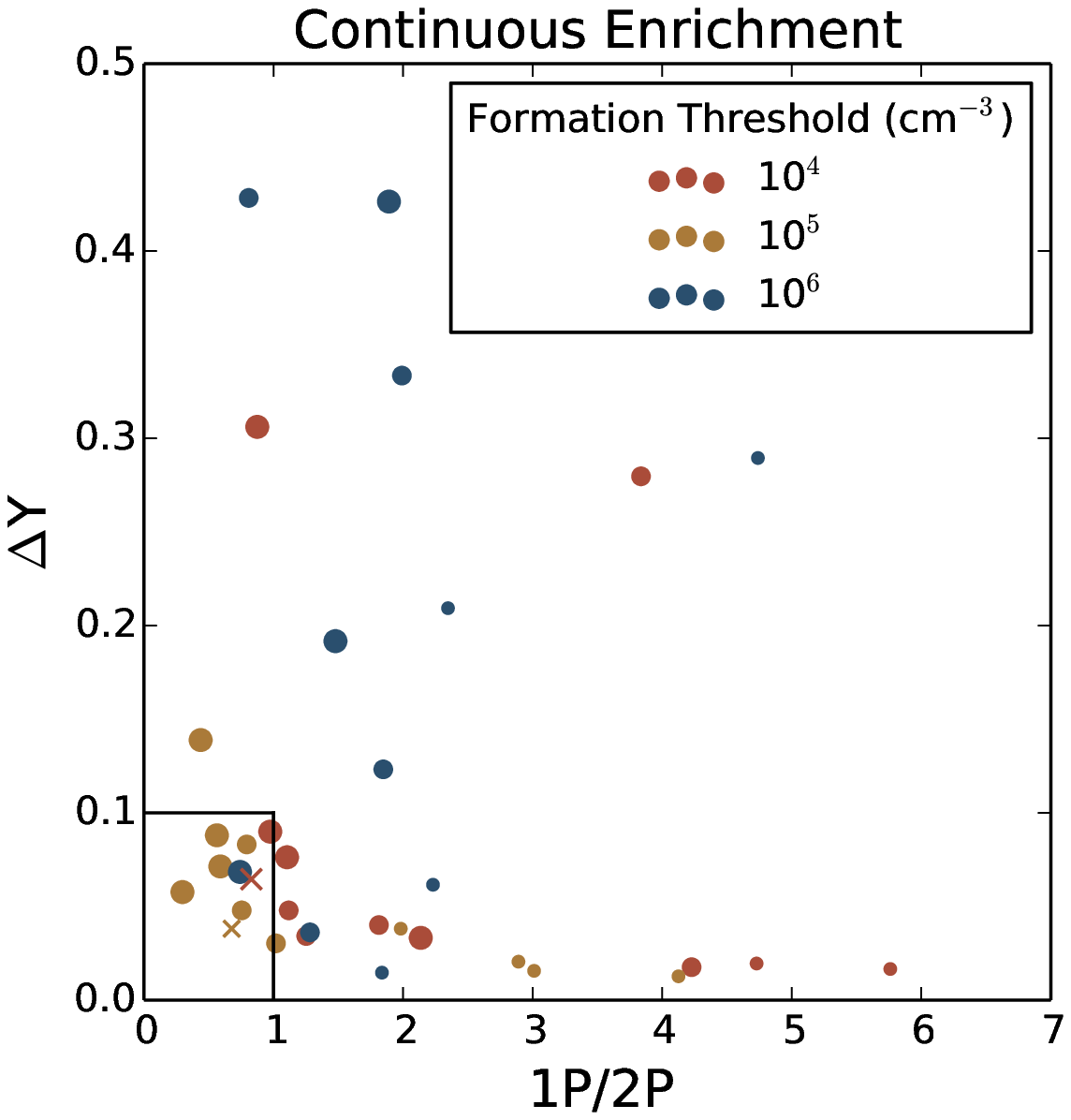}
\includegraphics[width=0.9\linewidth]{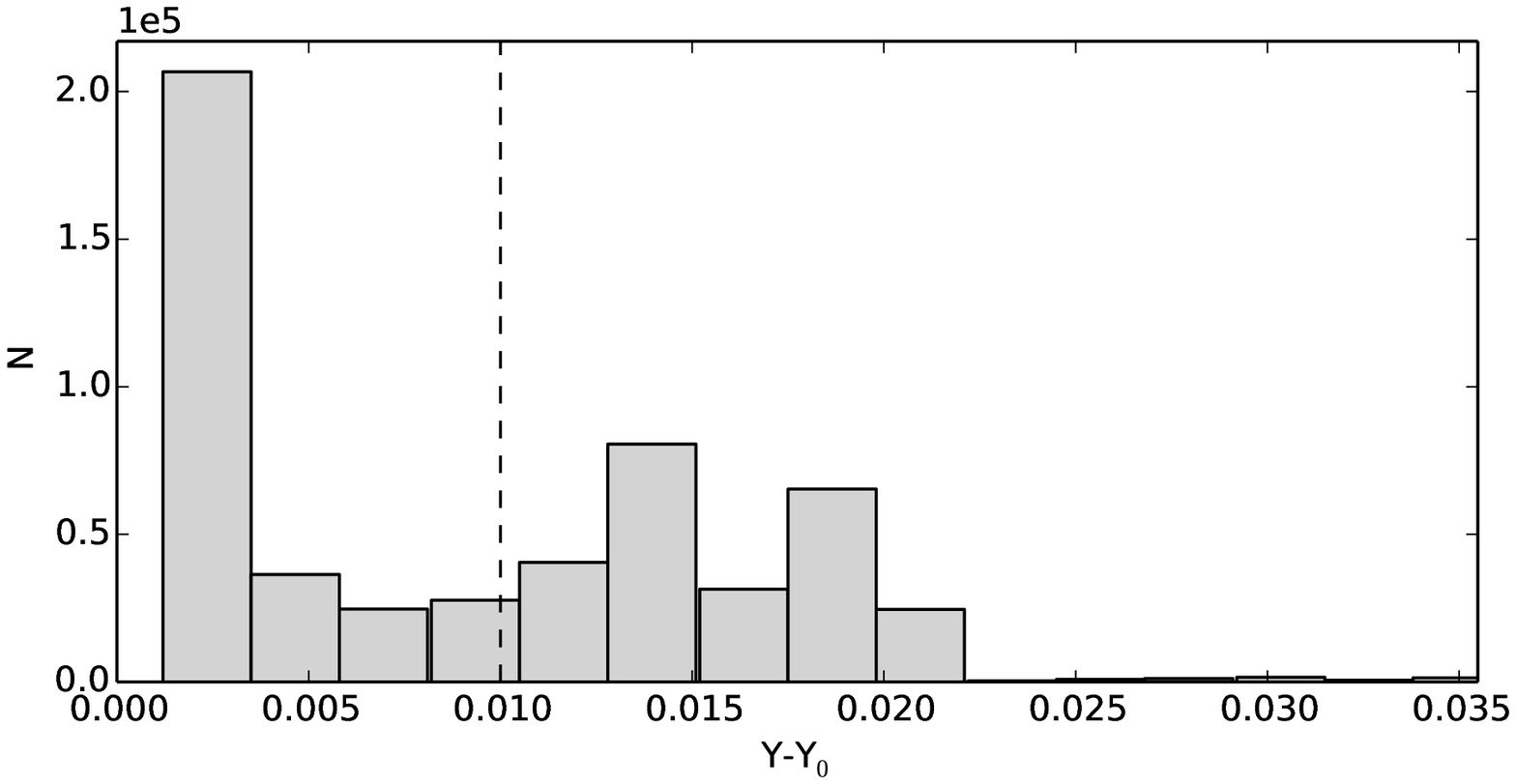}
\includegraphics[width=0.9\linewidth]{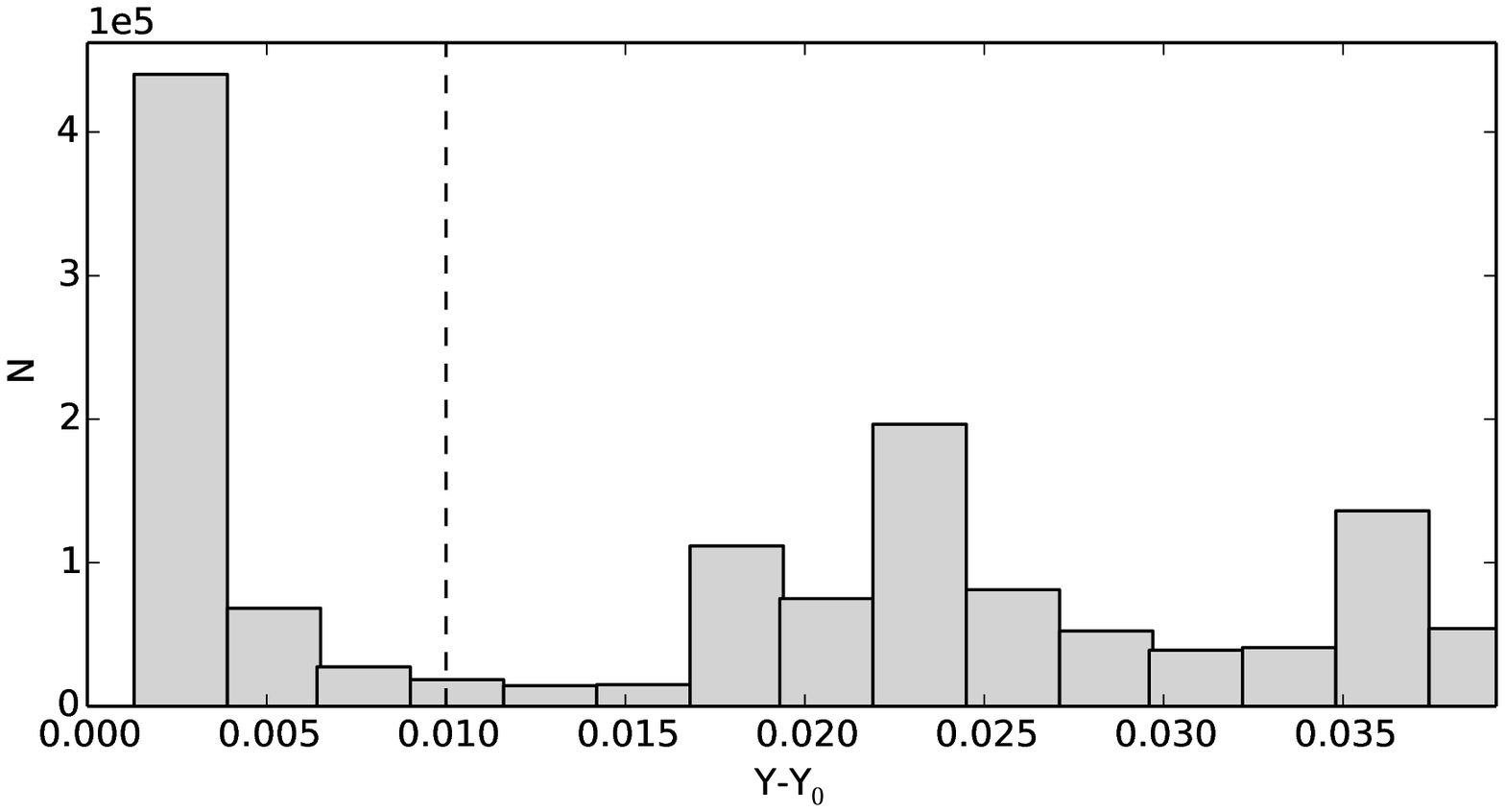}
\label{}
\caption{\textbf{Cluster abundance spreads and 1P/2P ratios for the continuous-enrichment model (CI).} 
	\emph{Top:} $\Delta Y$ against 1P/2P for individual star clusters. Larger filled circles
correspond to higher values of $f_{He}$.  Some models with extreme choices for
the parameters fall beyond the axis limits and are not shown here.  The black box represents the values 
consistent with observed GCs. \emph{Middle and Bottom:} Stellar $\Delta Y$ histograms for the two 
clusters marked by the crosses in the top plot. The vertical dashed line shows our adopted divide between 1P and 2P.}
\label{fig:Ostar}
\end{figure}

\begin{figure}
\includegraphics[width=.9\linewidth]{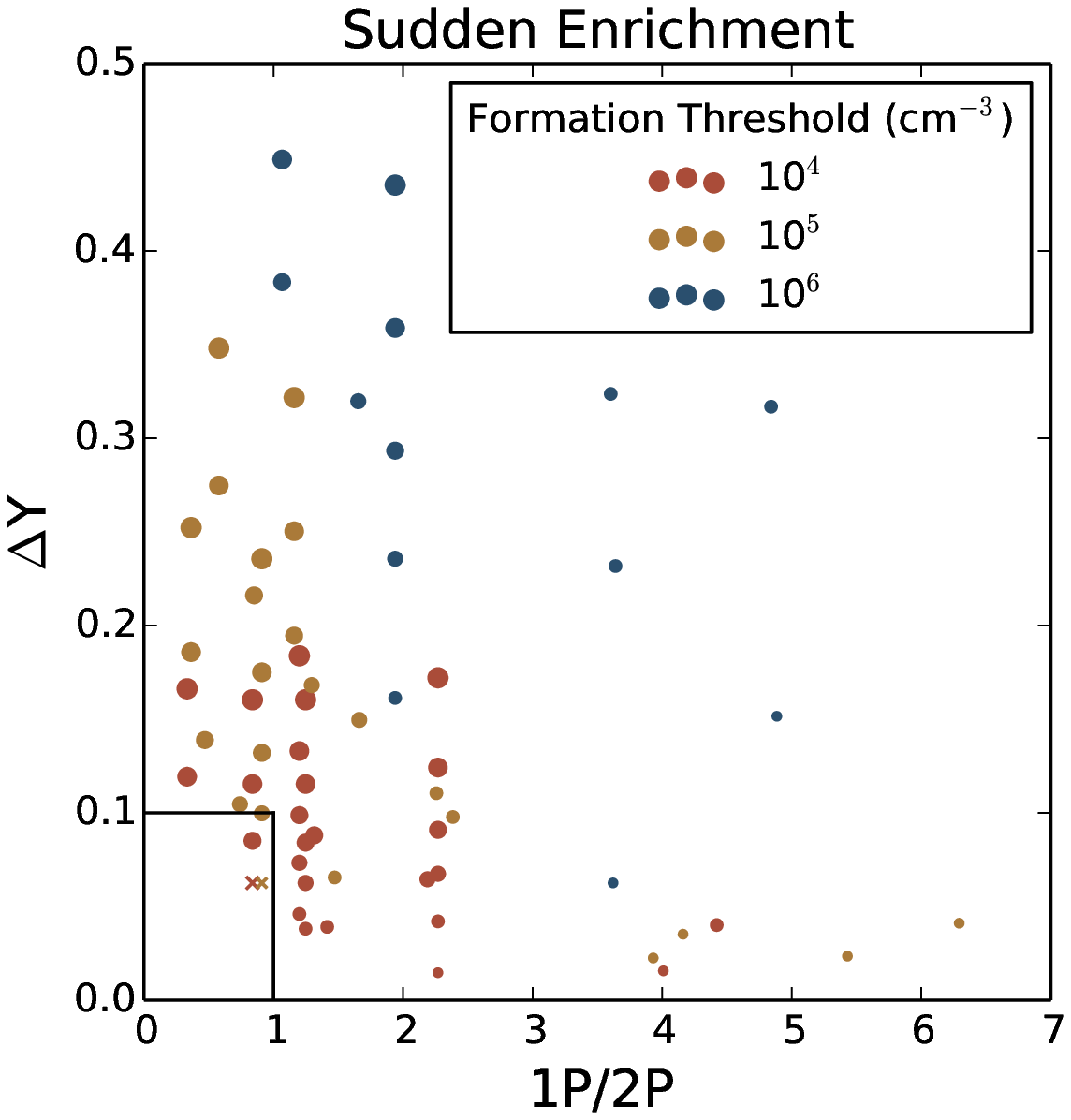}
\includegraphics[width=0.9\linewidth]{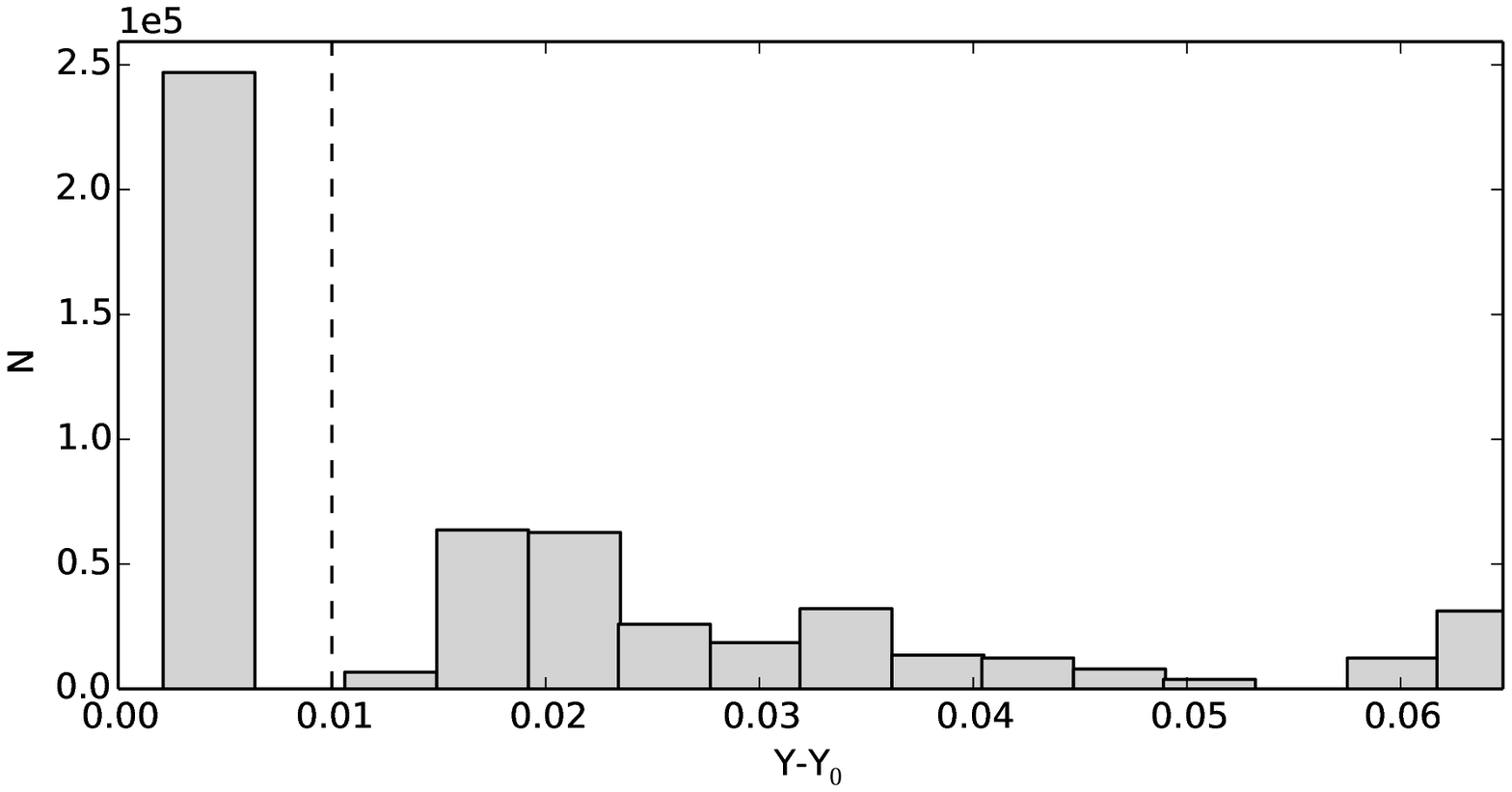}
\includegraphics[width=0.9\linewidth]{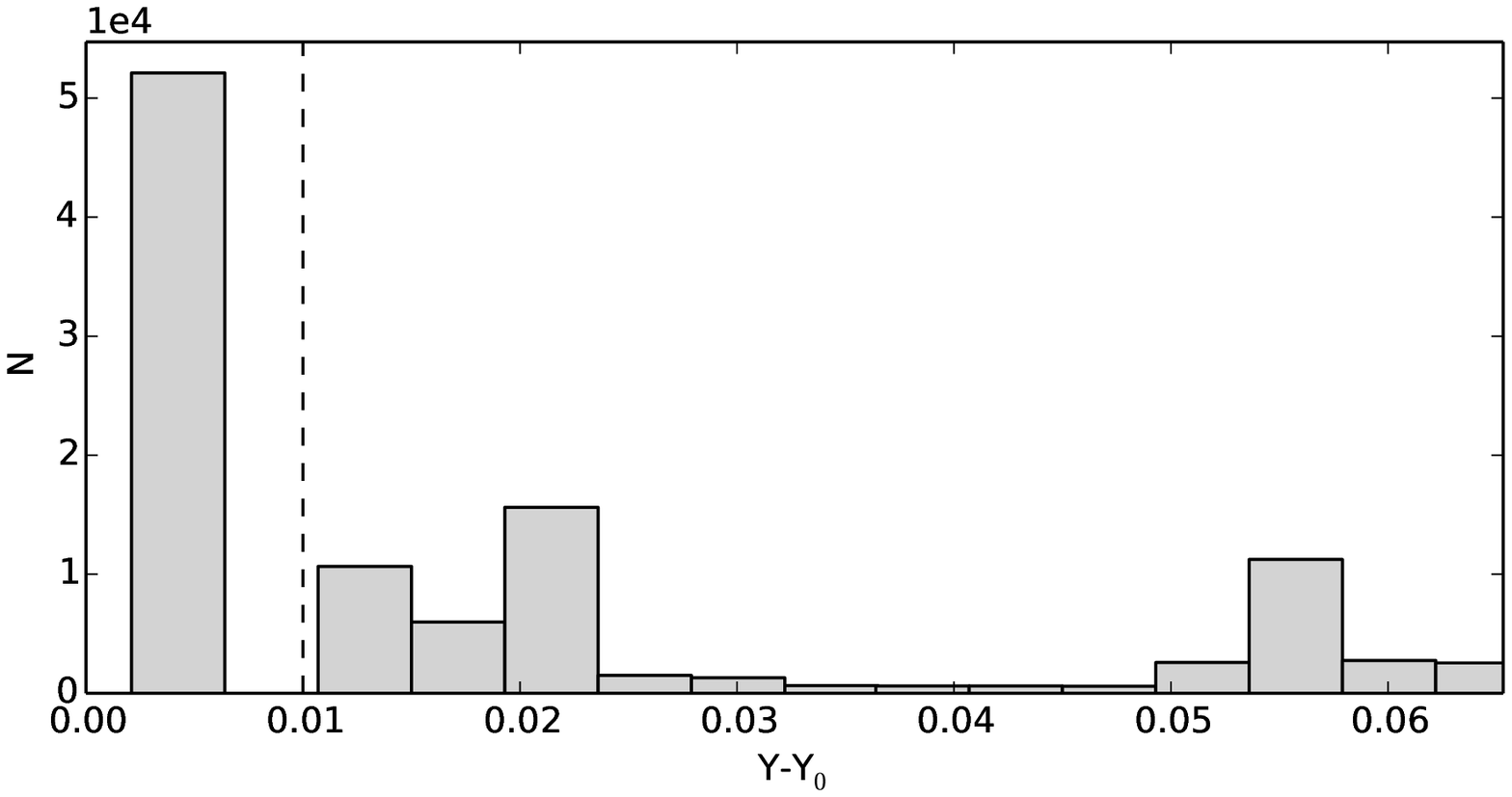}
\caption{\textbf{Cluster abundance spreads and 1P/2P ratios for the sudden-injection model (SI).} 
	\emph{Top:} $\Delta Y$ against 1P/2P for individual star clusters. Larger filled circles
correspond to higher values of M$_{He}$/M$_{cl}$. The black box represents the values 
consistent with observed GCs. \emph{Middle and Bottom:} $\Delta Y$  histograms for the two 
clusters marked by the crosses in the top plot. The vertical dashed line shows our adopted divide between 1P and 2P.}
\label{fig:SMS}
\end{figure}

In Figures \ref{fig:Ostar} and \ref{fig:SMS}, we show both $\Delta Y$ and the 1P/2P number ratio for individual simulation runs. In the left panels each point represents an individual model.  
For CI, 9 models fall within the observational range 
(solid black box) and of these, 6 have $f_{He} = 0.5$ 
while 3 have $f_{He} = 0.25$. Two examples of the stellar He abundance histograms appear below the summary figure.  In these histograms, distinct peaks are present but are bridged by fairly continuous $\Delta Y$ distributions. The first example in Figure \ref{fig:Ostar} (upper histogram) shows a relatively modest enrichment history reaching only to
$(Y-Y_0) = 0.02$, whereas the second example (lower histogram) generated three major peaks at $\Delta Y = 0.00, 0.02, 0.04$ but with some stars in between. 

The SI case generates $\Delta Y$ distributions with somewhat more distinct subpopulations. 
Four of the SMS models fall within the observational limits, all of which have formation thresholds of 10$^4$ or $10^5$ cm$^{-3}$. 
Two are illustrated in Figure \ref{fig:SMS}.  Furthermore, only clusters with M$_{He}$/M$_{cl}$ $=$ 0.03, 0.05, and 0.07 are consistent with observations, 
further constraining the valid range of injected masses of He.

An overall assessment of the model results can be fairly simply stated.  To produce a realistic level of $\Delta Y$ values, it is necessary to inject newly generated He (along with the accompanying light-element products of hot H burning) adding up to a few percent of the young cluster's mass (either gradually, or all at once).  The single-injection case (SI) is understandably an easier route to yielding a sharp bimodal $\Delta Y$ distribution.  But many of the observed MP cases are not simply bimodal \citep{Milone2017,Milone2018}, and
interestingly, either of the extreme cases (CI or SI) that we look at here are capable of yielding a wide variety of $\Delta Y$ distributions that resemble the observations.

\section{What Stars Are Producing the Internal Enrichment?}

A key question that has dominated the discussion of the MP phenomenon has been the physical source of the enriched elements.  A fundamental assumption in our model is that GCs are 
simply the high-mass extension of normal cluster formation.  In this view, the internal enrichment should therefore be coming from very massive stars that should be expected to form \emph{normally} under these high-mass, high-density conditions.  Such stars also need to be able to shed enriched material quickly, within $\sim 5$ Myr after the beginning of star formation and before SNe clear or heat the remaining gas.

For the continuous enrichment (CI) mechanism, we suggest O star binary (OSB) evolution \citep{deMink2009}, during which interactions between O stars in tight binaries shed enriched material that is then incorporated into nearby star-forming gas.  
Massive close-binary systems form rapidly in a nascent cluster and interactions between the stars shed a large amount of their mass which has been processed by hot hydrogen burning
\citep{deMink2009}. These ejecta have low velocities compared to radiatively driven winds from massive stars and are therefore efficiently trapped in the cluster. 
The processed material can then be incorporated into new, nearby protostars. Moreover, a significant fraction of the binary system's mass is lost during its evolution \citep{Petrovic2005}. The enrichment continuously increases as more and 
more OSBs form and die in the rapidly assembling cluster. This effect is likely further enhanced in massive, dense clusters where encounters with nearby stars will harden the short-period binary system, increasing the mass loss relative to field populations.

Calculations for a particular interacting binary (20+15 $M_{\odot}$ on an initial 12-day orbit) show that this object can reproduce most of the observed chemical abundance patterns in GCs including their He spread \citep{deMink2009}. Binaries of different masses and with different initial periods will produce different abundance patterns. 
Net yields integrated over an entire population of binaries are not yet available, but the expectation is that interacting OSBs can match the variety of abundance patterns seen in GCs \citep{BastianLardo}.

In our CI model, for simplicity we assume that every massive star $\geq$16 M$_{\odot}$ forms in an interacting binary system. Assuming a binary fraction of 100\% for massive stars necessarily places an upper limit on the amount of possible enrichment by 
this mechanism. However, estimates of the actual fraction of massive stars that are found in interacting binary systems are high, typically in the range 50-70\% \citep{Sana2012}. We then assume that each massive star sheds a fraction of its mass as He once it reaches a certain age. 
As described above, the subgrid model for star formation records the mass and formation time of each star that forms in every cluster, allowing the ages of massive stars to be determined. 

The sudden-enrichment (SI) case is more difficult to identify with a particular source (though as described above, the SI calculations were done primarily as a way to show the range of possible numerical outcomes in the problem).  
However, one possibility suggested in the previous literature is a SuperMassive Star (SMS) 
above $\sim 1000 M_{\odot}$ \citep{PZ2004,DeniHart,Gieles2018}, though such objects are still highly theoretical.  The main arguments in favor of SMSs at this stage are likely to be that (a) the extremely high gas densities in the centres of GC-scale protoclusters are the most likely places where SMSs could form; and that (b) stellar evolution models suggest that their high internal temperatures are in the right range 
to reproduce many of the right ratios of proton capture elements seen in GCs including C-N, Na-O, Mg-Al, 
and Na-F anticorrelations \citep{Sills2010, DeniHart}. The extreme mass of these objects also means that their lifetimes
should be exceedingly short, typically experiencing a supernova around 3 Myr after formation \citep{PZ2004}. There is considerable uncertainty regarding the fate of these objects, but their short lifetimes would manifest as a rapid injection of He and other elements into the surrounding cluster gas.

The first models of SMS formation were numerical N-body simulations demonstrating that stellar collisions in the centers of dense, 
gas-free clusters undergoing core collapse can lead to the formation of SMSs that are $\sim$1\% of the host cluster's mass in 
only 1-3 Myr \citep{PZ2004,Freitag}. More recent calculations, however, have shown that SMS formation can occur in massive, 
gas-rich clusters that are undergoing strong gas accretion \citep{Gieles2018}. In this framework, the inflowing gas mass 
of a cluster containing $\sim$10$^6$ stars and a half-mass stellar density $\rho_h > 10^3$ M$_{\odot}$/pc$^3$ causes the 
cluster to contract. The enhanced central density gives rise to an SMS with a mass of $\sim$10$^4$ M$_{\odot}$ in $<$5 Myr 
through both stellar collisions and gas accretion. The clusters formed in our simulations are consistent with these conditions. 
Moreover, as shown in our previous work \citep{NatAst}, these massive clusters have large gas accretion 
rates ($\sim$10$^5$ M$_{\odot}$/Myr) at these initial stages, further suggesting that they are potential locations for SMS formation.   

In our SI model, we assume simply that He injection occurs 2 Myr after cluster formation.
This time is likely a lower limit based on the numbers discussed above, but was deliberately chosen to provide the cluster with the most amount of time to form a second population of He enriched stars.  
The other parameter for the SI case is the total mass of He injected into the cluster gas reservoir, which equals $M_{SMS}$ times its mass fraction of He (assuming that it is completely destroyed at the end of its lifetime).  
The He mass fraction in a SMS can possibly reach as high as $Y = 0.40$ \citep{DeniHart}.  
Moreover, some of the star's mass would be retained by the post-supernovae remnant black hole, whose expected mass is quite uncertain. 

In short, the SMS assumption still has major unknowns.  Again for numerical simplicity, our SI calculations assume $Y=1$ or $M_{SMS} = M_{He}$, which effectively places a strong lower limit on the true SMS mass.  
For the calculations we vary the mass ratio
$M_{He}/M_{cl}$ from 0.01 to 0.15, which corresponds to SMS masses of $\sim1000-50,000 M_{\odot}$. We note that 
a mass ratio of 0.01 is more in line with the original N-body simulations \citep{PZ2004}, but those works do not consider 
the presence of gas in the cluster which may act to increase the mass of the SMS. 
Our simulations with a high-density formation 
threshold have a higher star formation rate. Since the amount of enrichment in this scenario is directly proportional to the 
total stellar mass formed within 2 Myr, clusters with a higher star formation rate will have higher SMS masses.

\section{Discussion and Summary}

Overall, our simulations indicate that sudden-injection enrichment generally produces higher $\Delta Y$ and more discrete abundance 
distributions.  On the other hand, gradual injection yields abundance distributions that are somewhat more continuously populated and with smaller $\Delta Y$.  As discussed above, the SI route is less easily linked with a convincing or well understood stellar source.
It is encouraging, however, that
both routes are capable of producing realistic 1P/2P ratios, as well as abundance distributions for the 2P (enriched)
population that can be complex and varied.  

In many real GCs, just two clearly distinct populations appear, as shown
in the chromosome maps for the survey of 57 Milky Way GCs \citep{Milone2017,Milone2018}.  But it is already clear that this is
not a universal outcome.  For many other GCs in the observational  
survey, 3 or more subpopulations are evident, and even cases of rather continuous abundance distributions with
no clearly identifiable `gaps', within the limits set by the observational uncertainties.  The enrichment model presented here is capable of producing this range of outcomes.

In both cases our model allows us to place firm constraints 
on the total amount of He required to reproduce the range of observed stellar He spreads and population ratios. It can easily be generalized 
to models of other stellar processes that are characterized by injection of large quantities of He into star-forming gas over finite but short time intervals.

The OSB approach would require a large fraction (25 - 50\%) of each O star's mass to be released as He, and 
it leaves behind $\Delta Y$ distributions that are usually not sharply bimodal.
In the SI calculation, the results we have so far indicate that the (postulated) SMS at the centre of the cluster needs to be at least 3-7\% of the cluster's mass, 
corresponding to $\sim$1000-10,000 M$_{\odot}$ in a 10$^{5-6}$ M$_{\odot}$ cluster.  
Observations that can better quantify 
the He abundance distributions will be able to distinguish more strongly the different enrichment paths.

In this paper, we have outlined a preliminary investigation of one particular route to producing multiple
stellar populations within massive star clusters.  At this point it is worth summarizing what we believe to be the major
advantages of the approach:
\begin{enumerate}
\item{} Perhaps most importantly, this interpretation of MPs is built on a rigorous, quantitative RHD model for star cluster
formation within GMCs.  MPs are seen as emerging as a direct byproduct of normal cluster formation without supposing that GC formation occurred in a fundamentally different way than does lower-mass cluster formation.
\item{} Both the 1P (pristine) and 2P (enriched) populations form actively and simultaneously with the first star formation within
young massive clusters, explaining in a natural way why no detectable age difference between them should be seen in real GCs.  
\item{} The classic ``mass budget" problem is essentially avoided entirely because clusters are built within GMCs rather than starting as isolated monolithic gas clouds. Here, the entire GMC provides the much bigger reservoir of gas that
a young cluster in formation can draw from through inflow along gaseous filaments.
\item{} The complex assembly of clusters within GMCs has an in-built stochasticity, which means that different clusters can 
experience radically different growth histories \citep{NatAst} as expected in turbulent, filamentary clouds.  Large differences in final chemical abundance patterns can therefore 
emerge between clusters quite naturally, but this variety itself is an important point of agreement with observations 
\citep{Renzini, Piotto, Milone2017}. The large cluster-to-cluster variety in the 1P/2P ratio and the degree of enrichment ($\Delta Y$) 
are also natural outcomes of this general model.
\item{} More massive clusters are better able to hold on to their gas reservoirs during formation, so the relative numbers
of 2P stars should increase with cluster mass, consistent with observations \citep{Milone2017}.
\item{} Even in this simple form, the model is quantitative enough to constrain either the numbers and ejected He mass fractions of O-star binaries or (much more speculatively) the mass range of SMSs.
\end{enumerate}

The modelling route presented in this paper explores two opposite limiting cases (smooth, continuous enrichment versus a large one-time event).  We regard this work as a first 
promising step that can be developed further in a number of ways.
For example, in practice young massive star clusters will certainly hold a significant population of O stars in close
binaries, but an SMS could also be present, so a combination of the two mechanisms could
be explored.  A true SMS may shed mass more continuously through its short lifetime \citep{Gieles2018}, so a delta-function 
pulse is certainly simplistic.  The true enrichment rate ($dY/dt$) need not be either artificially smooth nor a delta-function,
given the stochastic nature of the gas inflow, merging with other protoclusters, and random sampling of the IMF.
Further work will also need to be done to compute more detailed abundance patterns of the light elements 
that are represented here only by the He enrichment.  

RHD modelling of the young clusters also needs to be
continued past 5 Myr through the supernova stage to track their survival. In addition, the model
GMCs have relatively high star formation efficiencies \citep{NatAst} that might be reduced to more realistic
levels by including several additional processes (winds, supernovae, and magnetic fields).  This step may in turn reduce the net $\Delta Y$ yields, which would bring 
more of the model clusters into the observationally valid range.
Finally, on the computational frontiers, new techniques may allow RHD simulations to be performed at sufficient resolution to resolve individual star formation in the young clusters themselves.  This will open up new capabilities in understanding what happens to the gas reservoir held by a protocluster and ultimately how realistic our general model can be.

\section*{Acknowledgements}

We thank an anonymous referee for a useful report.   Computations were performed on the GPC supercomputer at the SciNet HPC Consortium and supported by ComputeCanada by a Resources for Research Groups (RRG) Grant to REP.  SciNet is funded by the Canada Foundation for Innovation under the auspices of Compute Canada; Government of Ontario; the Ontario Research Fund: Research Excellence; and the University of Toronto.  The authors acknowledge financial support from the Natural Sciences and Engineering Research Council of Canada (NSERC).  




\bibliographystyle{mnras}
\bibliography{He_paper}


\bsp	
\label{lastpage}
\end{document}